\documentclass[11pt,dvips]{article}
\usepackage{amssymb,amsmath}
\newcommand{\beg}[2]{\begin{equation}\label{#1}#2\end{equation}}
\newcommand{\rref}[1]{(\ref{#1})}
\newcommand{\cform}[3]{\begin{array}{c}
{\scriptstyle #3}\\
#1\\
{\scriptstyle #2}\end{array}}

\author{Igor Kriz\thanks{The author is supported by NSF grant DMS 0305853}}
\title{Some remarks on fundamental physical F-theory}

\newcommand{\R}{\mathbb R}
\newcommand{\C}{\mathbb C}

\newcommand{\Z}{\mathbb Z}

\begin{document}

\maketitle

\begin{abstract}
The purpose of this paper is to investigate the possibility of
a physical 12-dimensional F-theory. We study the question of
geometric interaction terms in the F-theory Lagrangians.
We also introduce a new supergravity multiplet in 
dimension $(9,3)$ which is based on a particle with $3$-dimensional
timelike worldvolume. A construction of signature $(9,3)$ 
F-theory is given using dualities analogous to those considered
by Hull, and possible matches of F-theory's low energy fields with
the $(9,3)$-supergravity field content is given. Finally,
preliminary suggestions are made regarding a possible
phenomenological compactificaton of F-theory 
from dimension $(9,3)$ to $(3,1)$.
\end{abstract}

\section{Introduction}

\vspace{3mm}
The purpose of this paper is to introduce some new evidence for a fundamental
physical $12$-dimensional
F-theory, and to exhibit, at least on a preliminary basis,
some of its implications. F-theory has been first suggested by
Vafa \cite{Vafa}, and has been subsequently 
mentioned many times in the literature (cf. \cite{Sag, Mstar, Sen, FMW}). 
In some cases, it has been merely suggested as a bookkeeping tool for
action terms in lower-dimensional compactifications, in particular
in the case of a fiber bundle with elliptic fiber over a type
IIB spacetime. Although this
suggests that more should be going on, and that a fundamental F-theory
unifying these contexts
should exist, there are difficulties with such a proposal. For example, 
it is not clear what the theory's low energy limit should be,
since supergravities in the classical sense do not exist
above dimension $11$.

\vspace{3mm}

The purpose of this paper is to investigate these
questions and to show new
indications that a fundamental F-theory, in fact, does exist, while
exploring how that would be possible and what it would imply.
The first motivation of this project was a previous joint investigation with
Sati \cite{KS,KSB,KSmod}, where we discovered evidence that
F-theory is relevant in a different setting, namely M-theory and
type IIA string theory. There, one doesn't have an elliptic fibration
over $10$-dimensional
spacetime, but a $12$-dimensional manifold with boundary which is
a spin cobordism of the spacetime of M-theory compactified on $S^1$.
Witten \cite{flux, DMW} found that this $12$-manifold must be considered
when investigating the effective action of M-theory. In \cite{DMW}, Diaconescu,
Moore and Witten compared the partition function of M-theory thus obtained
with the partition function of type IIA string theory which is calculated
using quantization by $K$-theory. Yet, it did not yet follow from this 
why F-theory action should be considered on this cobordism manifold,
and how it relates to the type IIB context. In particular, in the IIB context,
certain action terms occur which did not have an obvious analogue in M-theory.
This concerns, in particular, the lift of the IIB field $G_5$.

\vspace{3mm}
Jointly with Sati, we more or less accidentally found an 
extension of the $K$-theoretical
partition function \cite{Wi1,DMW} of type II string
theory, which uses elliptic cohomology
instead of $K$-theory. The first indication
that such function
should exist was
just the observation \cite{KS} that an anomaly $W_7(X^{10})$ found by \cite{DMW}
is in fact the same as the obstruction to orientability 
of spacetime $X^{10}$ with respect to
elliptic cohomology. When actually defining the function however, 
one needs to define a ``quadratic structure''
(which amounts to a real structure),
at which point other obstructions emerge, notably $w_4$ or $\lambda$,
(depending on whether one uses real-oriented elliptic cohomology
or topological modular forms).
This suggests that this elliptic partition function 
has some connection to type I and heterotic
string theory, where one sees such anomalies.
Trying to interpret this function, we see that it
indeed arises as free
field approximation of a certain dynamical theory. We eventually concluded
that the theory needed is F-theory, and proposed that an analog of
the investigation of \cite{DMW}, which would match the elliptically refined
partition function of type IIA string theory with the partition function
of F-theory, should be possible using Witten's loop version \cite{wi}
of the Dirac operator. This proposal was recently partially carried out by
the author and Xing \cite{KX}.

\vspace{3mm}
In the case of type IIB string theory the evidence for elliptic partition
function is
even stronger. It was noted in \cite{DMW} that a twisted $K$-theory
field strength which is the natural extension of the $K$-theory field
strength considered there (see also \cite{Wi1}) violates S-duality
of type IIB. The author and Sati \cite{KSB, KSmod} in fact showed
that this cannot be remedied simply by modifying the definition
of twisted K-theory without introducing other fields. Their investigation
suggested that the problem is in the twisting itself, and that to remedy
it, one needs a theory where the field $H_3$ doesn't introduce
a twisting. The natural candidate is again
elliptic cohomology (or more precisely the theory of topological
modular forms) where $H_3$ ``untwists'' and merely is represented
by multiplication
by some element of that generalized cohomology
theory. Again, it is a natural suggestion that this partition
function should match the partition function of 
F-theory is a theory on $12$-dimensional spacetime (although
at the present time, we don't have a proof yet). The type of F-theory
which arises in this case is on $Z^{12}=X^{10}\times E$ where
$E$ is an elliptic curve. Modularity of our elliptic partition function
is, by the proposal of \cite{KSmod}, related to modularity of 
the first cohomology group $H^1(E)$. We further proposed that modularity
of the elliptic partition function contains S-duality in type IIB string
theory.

\vspace{3mm}
However, how are the F-theories arising in types IIA and IIB related?
How would one get a hold of the relevant terms of its action? And
what is its low energy limit? These questions are the main subject
of the present paper. Actually, a beginning of the
discussion of the action is 
in \cite{KX}. There, we first of all explained how it is possible that the
$Z^{12}$ which is a spin cobordism of M-theory spacetime only seems
to have a $G_4$-field (a lift of the $4$-dimensional field strength
in M-theory, see \cite{flux}), while the action of the F-theory 
which is elliptically fibered
on IIB seems to involve a $G_5$-field. When investigating the $12$-dimensional
theory which could have the same partition function as the elliptic
refinement of IIA, the natural object to look at was a $G_4$-field on
the {\em free loop space} $LZ^{12}$, which can be investigated using
Witten's formula \cite{wi} for index on loop space. But a $G_4$-field
on $LZ^{12}$ produces both a $G_4$ and $G_5$ on $Z^{12}$. In \cite{KX},
we were interested mostly in the case when this field is pulled back
from $G_4$ on $Z^{12}$, in which case we obtain a refinement of the
M-theory partition function, which, indeed, turns out to match, in an
appropriate sense, the elliptic refinement of the IIA partition function.
That was the main result of \cite{KX}. In the present paper, we
investigate the general case, in which the relevant interaction term
can be described as $E_8$-index on the loop space $LZ^{12}$. In particular,
we find that there is a quantization condition on $G_4$ and $G_5$ that they
must jointly lift to an $E_8$ bundle on $LZ^{12}$, but except that
condition, $G_4$ and $G_5$ are in fact independent fields in F-theory,
i.e. there is no explicit constraint between them.

\vspace{3mm}
Next thing we considered in \cite{KX} was how the F-theory on a cobordism
of M-theory spacetime and F-theory elliptically fibered on IIB spacetime
could be related by duality. We proposed, in fact, such a process: in
the case of IIA spacetime $X^9\times S^1$ subject to T-duality, we obtain
M-theory on $X^9\times S^1\times S^1$. Given appropriate spin structure,
we can obtain a spin cobordism $Z^{12}$ which essentially involves only
the $S^1\times S^1$ coordinates. Now the appropriate analogue of T-duality 
shrinks only the boundary (the M-theory spacetime) to a space one dimension
lower (by contracting one copy of $S^1$ to a point). This corresponds to turning
$Z^{12}$ to a manifold without boundary. In a suitable case, this is, in fact,
the ``T-dual'' F-theory fibered over type IIB spacetime. There is, however,
by similar arguments, also
another type of duality which is
a self-duality of F-theory without boundary.
This duality interchanges the $G_4$ and
$G_5$ fields. This is what makes it feasible to conjecture that there is
only one universal F-theory rather than separate theories of ``types IIA and IIB'',
and possibly other types.

\vspace{3mm}
Now it remains to explain how this would work in physical signatures.
We address that in the present paper. Hull \cite{Mstar} 
considers a $\text{IIA}^*$-theory, which is a T-dual of IIB theory on
a timelike coordinate of signature $(9,1)$ spacetime. The strong coupling
limit is $\text{M}^*$-theory, which has spacetime of signature $(9,2)$.
The cobordism one gets in the last paragraph in that case has signature
$(9,3)$. Thus, applying the duality described above, one can construct
F-theory of signature $(9,3)$.\footnote{The possibility of such theory
is in fact mentioned in \cite{Mstar}.}

\vspace{3mm}
The next interesting question is what the low energy limit of such theory
might be, given the fact that there is no $12$-dimensional supergravity
in the conventional sense. We give a proposal for that in the present paper:
in a truly physical spacetime of signature $(9,3)$, particles should have
$3$-dimensional worldvolumes with three time-like coordinates.
Looking at massless spectra of such type, we get different supersymmetry
conditions. In fact, in signature $(9,3)$, we are looking for a supermultiplet
which is a representation of a supergroup extension of $Spin(6)$.
In this paper, we find in fact a candidate for such supermultiplet which
seems to match the fields of F-theory, and thus could be its low energy limit.

\vspace{3mm}
Supposing, finally, that this proposal in fact works, 
certain interesting possibilities arise. In particular, there is a new
type of compactification from dimension $(9,3)$ to dimension $(3,1)$
based on the fact that $(9,3)=3\cdot(3,1)$. At the end of this paper,
we briefly speculate what this could mean for phenomenology. It is possible
that instead of splitting spacetime as a direct sum where excess dimensions
are ``discarded'', a $(9,3)$-dimensional space could in fact be represented
as a tensor product, where one factor is phenomenological spacetime.
Confinement using the $3$-brane in F-theory wrapped around a topologically
non-trivial submanifold could explain why in the infrared, measurements
are approximated by $(3,1)$-dimensional measurements, while however
all $12$ dimensions are in fact functional as spacetime dimensions.
Such theory may have potentially interesting applications, including,
for example
a better explanation of distance decay of strong interaction.

\vspace{3mm}
The present paper is organized as follows: In section \ref{lag} below,
we shall discuss the action of F-theory, in particular the geometric
term which we arrived at as a result of the investigation in \cite{KX}.
In section \ref{susy}, we shall discuss the supermultiplet
candidate for signature $(9,3)$ supergravity based on a massless particle with
three world volume time dimensions. In section \ref{dual}, we review the
F-theory duality discussion from \cite{KX}, and apply it to the case of
physical signatures following the method of Hull \cite{Mstar}, constructing
an F-theory which could produce the field content predicted by the
analysis of section \ref{susy}. In section
\ref{phen}, we present the phenomenological applications which could
arise from the present theory. Section \ref{egen} contains some concluding
remarks.

\vspace{3mm}
{\bf Acknowledgement:} I am indebted to H.Sati and H.Xing for collaborations
on F-theory 
in the context of the IIA and IIB compactifications.

\section{On the action of F-theory}
\label{lag}

Let us now work 
to describe the action of F-theory in terms of the potentials $A_3$, $A_4$,
and corresponding field strengths $G_4$, $G_5$. There will be, of course, a 
standard kinetic term of the form
\beg{est}{\int_{Z^{12}}G_4\wedge *G_4 +G_5\wedge *G_5,}
but there are other interaction terms, including topological and geometric
terms which we need to identify.

\vspace{3mm}
We begin by recapitulating known facts. First, let us
recall the analysis of the topological Chern-Simons term in M-theory,
as analyzed in \cite{flux,DMW}. There, we look at an $11$-dimensional
manifold $Y^{11}$. Its M-theory Chern-Simons term should be a topological
term, and should be expressed by finding a spin $12$-manifold $Z^{12}$
whose boundary (with spin) is $Y^{11}$. The leading term is to be
\beg{elead}{\frac{1}{6}\int_{Z^{12}}G_4\wedge G_4\wedge G_4, 
}
but one must adjust the expression so that it doesn't depend on
the choice of the spin cobordism $Z^{12}$. In particular, we want
to get $0$ on a closed $12$-manifold $Z^{12}$.
It turns out (\cite{flux}) that the corrected expression is
\beg{ecobord}{\frac{1}{6}\int_{Z^{12}}G_4\wedge G_4\wedge G_4 +G_4\wedge I_8,
}
with $I_8=(p_2-\lambda^2)/48$. 
The second term of \rref{ecobord} can
be interpreted as a $1$-loop gravity correction term. When calculating
the partition function of M-theory, \rref{ecobord} actually contributes
the phase.
This raised the first provocative question to the author and
Sati that a $12$-dimensional theory may be relevant (although, of course,
F-theory was originally proposed in a different context by \cite{Vafa}).
Now the action \rref{ecobord} cannot be the whole story for F-theory $Z^{12}$,
precisely due
to the fact the corresponding phase factor makes sense on a manifold
with boundary, and vanishes on a closed manifold. In \cite{Vafa}, on the
other hand, one considers closed manifolds only, which are in
fact bundles
on type IIB spacetime with fiber an elliptic curve (see also 
\cite{FMW,Sen}).

\vspace{3mm}
In \cite{KSmod}, jointly with
H. Sati, we showed that
in fact very likely considering
type IIB string theory as F-theory with an elliptic curve
fiber is necessary for
the consistency of S-duality in type IIB string theory in the
presence of topological terms. (The breaking of S-duality
on the level of twisted K-theory approximation was actually 
first noted in \cite{DMW}; in \cite{KSB}, we showed that
this cannot be fixed simply by modifying somehow the
definition of twisted K-theory in the presence of the same fields.)
In the F-theory setting, S-duality is 
recovered via a relation between S-duality
and modularity in $H^1(E)$ where $E$ is the fiber. Very interestingly,
similar relations in fact also emerged much earlier in Borcherds-Harvey-Moore theory,
\cite{bor,mh,mh1}. Sati and the author plan to pursue this connection in future work.

\vspace{3mm}
In fact, what we did in \cite{KS,KSB} 
was construct a deformation of the K-theory partition
function for type IIB and IIA string
theory which is modular, using elliptic cohomology instead of K-theory,
and conjectured (which was later partially confirmed in \cite{KX}
in the case of IIA), that
this is equal to the partition function of F-theory (for those
phases for which the F-theory action indeed reduces to IIA).
In fact, to be precise,
getting exactly the right function requires the use of the 
Hopkins-Miller theory
of topological modular forms $TMF$ (see \cite{KX}). We shall
return to that point below.

\vspace{3mm}
For now, however, let us remark that
while exploring
that connection, another piece of the picture emerged: to define the
elliptic partition function,
a $4$-dimensional obstruction showed up, which is of a similar nature
as obstructions in type I and heterotic string theory. This was our
first suggestion that there should be a fundamental
F-theory which would indeed unify all 10-dimensional string theories. 

\vspace{3mm}
Let us now recall what was known about the action of F-theory
with an elliptic curve fiber.
Following essentially the idea of \cite{Sag}
who worked in the case of Calabi-Yau compactifications, 
we proposed in \cite{KSmod}, in the context of
F-theory compactified on an elliptic curve, the term
\beg{ev12}{\frac{1}{6}\int_{Z^{12}}A_4\wedge G_4\wedge G_4+ A_4\wedge I_8
}
where $A_4$ is a $4$-form potential, $G_4$ is $4$-form
of M-theory lifted to $12$ dimensions
and $I_8=(p_2-\lambda^2)/48$. Specifically, \cite{Sag} investigated
subharmonic expansions of particular form of the potentials $A_3$, $A_4$ on 
$Z^{12}=M^6\times CY$ and showed that then the leading
term of \rref{ev12} recovers
the expected $6$-dimensional coupling term. They also note that
for similar reasons, the terms
\rref{ecobord} must be present.

\vspace{3mm}
There are, however, questions about the formula \rref{ev12}.
First of all, \cite{Sag} note that in their
expansion, the potentials $A_3$ and
$A_4$ are not independent.
In other words, they conjectured that there must
be another independent relation between $A_4$ and $G_4$. Yet, as
far as the author knows, such equation has never been found. In fact, below
we will exhibit a curious form of signature $(9,3)$ SUGRA whose fundamental
object is a massless particle with $3$-dimensional lightlike world volume 
(i.e. a particle moving at the speed of light in three independent
time dimensions, which match the $3$ spacetime time dimensions).
In that setting, unusual field contents emerge. In particular, in
lightcone gauge, we get representations of the group $Spin(6)$.
There, we will see that candidates for both $A_3$
and
$A_4$ occur and are independent (although curiously, $A_3$ is in the
same representation as the graviton, which leads us to speculate that
$A_3$ and the graviton merge). 

\vspace{3mm}
This is evidence that $G_4$, $A_4$ indeed should be independent. It is possible
that the relation suggested in \cite{Sag} occurs when looking at a particular
sector of the theory. It is not unusual that when looking at compactifications,
fields endowed with additional constraints explain terms in the compactified
theory: for example, in \cite{DMW}, to get type IIA partition function from
M theory on $X\times S^1$, one assumes that $G_4$ is pulled back from $X$.
Nevertheless, we will see below that there is in fact a tie between $G_4$ and
$G_5$, although not exactly in the form of an equation.

\vspace{3mm}
We made further progress on answering these questions in \cite{KX}.
The purpose of that paper was to derive an analogue of \cite{DMW}, which
would match the elliptic partition function on $X^{10}$ of type IIA with
a partition function of an F-theory. Thus, this required defining F-theory
beyond the case of an elliptic fibration on IIB.
Following a proposal in \cite{KSmod}, the approach we
took in \cite{KX} was to look at loop versions of the sum of indexes 
\beg{ew1}{\frac{I_{E_8}}{2}+\frac{I_{RS}}{4}
}
considered
in \cite{DMW}, which is equal to \rref{ecobord}. Here \rref{ew1} means
indices of the Dirac operator twisted by the adjoint $E_8$-bundle and
the shifted complexified tangent bundle, respectively. The $E_8$-bundle
is associated with the $4$-dimensional integral cohomology class corresponding
to $G_4$.

\vspace{3mm}
By `loop version', we mean the following. The class $G_4$ can be pulled
back to the loop space $LZ^{12}$. Witten \cite{wi} found a way of
defining the Dirac operator on loop space. Working in this way
requires the condition
\beg{ew0}{\lambda{Z}=0,
}
which in fact implies vanishing of all the $4$-dimensional anomalies mentioned
above.
Although this has not been made mathematically
rigorous yet, the index is a space with an $S^1$-action, and the 
trace of $q\in \C^{\times}$ can be computed using a trace formula,
thus giving a power series in $q$, which is a modular form given the
condition \rref{ew0}. The pullback of the class $G_4$ to $LZ^{12}$ can
be considered $S^1$-equivariant in a neighborhood of fixed loops, so
we get a version of $I_{E_8}$ which is a modular form. This is what
we did in \cite{KX}. (To be completely accurate, we essentially
neglected the Rarita-Schwinger term, and replaced its treatment by
just dropping a summand of the loop index which contains $p_3$.
This is justified in a first approach, as one expects a more complicated
boundary behavior in loop spaces, so the exact Ho\v{r}ava-Witten
analysis of anomaly cancellation will be more difficult to carry out.)
In summary, the $E_8$-index on loop space $LZ^{12}$
(in the case of field strength pulled back from $Z^{12}$)
gives the formula
\beg{ef3}{\int_{Z}G(\frac{1}{6}G^2-5S_4p_2)
}
where $S_{\ell}$ is the Eisenstein series
\beg{ef2e}{S_{\ell}=S_{\ell}(\tau)=-\frac{B_{\ell}}{2{\ell}} + 
\cform{\sum}{n=1}{\infty}\left(\cform{\sum}{d|n}{}d^{\ell-1}
\right)q^n.
}

\vspace{3mm}
Now we matched in \cite{KX} the resulting `phase term' with the
phase factor of a
variant of the elliptic partition function proposed in \cite{KS} on $X^{10}$ of
type IIA (where, as above, $Z^{12}$ is a spin cobordism of $X\times S^{1}_{R}$).
But let us first consider what we computed. To visualise the situation,
we can imagine that $G_4$ is the field strength associated with a 
$2$-brane $M$, i.e. a $3$-dimensional world volume. But now we have considered
the propagation of this $2$-brane $M$ on $LZ^{12}$, which is equivalent
to the propagation of $M\times S^1$ on $Z^{12}$. We see that therefore
we are in fact describing the propagation of a $3$-brane, which corresponds
to a potential $A_4$ or field strength $G_5$.

\vspace{3mm}
Other observations however must be made. First, the role of $q$ is that
the brane, which has one $S^1$-factor, propagates along another copy of
$S^1$, and we take the
trace. The $q=e^{2\pi i\tau}$ measures the moduli parameter of
the elliptic curve $E$ which is the product of these two copies of $S^1$.
Thus, the first observation is that this index is not a topological
invariant. It appears, however, reasonable to conjecture that it be
a {\em conformal} invariant of the brane, since it is a product
of a topological invariant and a conformal invariant.

\vspace{3mm}
Next, we can in fact only speak of a ``phase'' if $q$ is in fact real, i.e.
the two directions are orthogonal. Otherwise, the factor contributes to
amplitude as well as phase. In fact, to even state the comparison with
the elliptic partition function in IIA, we must assume that the moduli
parameter has special values, where the elliptic curve is defined by
an algebraic equation with integral coefficients: in that case, the
phase is in fact a topological invariant and (as follows from the
results of \cite{KX}), doesn't depend on the choice of $Z^{12}$). This
restriction to special values is not completely unexpected, it is 
a similar effect as e.g. the behavior of special values of L-functions.

\vspace{3mm}
Let us return however to the problem of the interaction term of F-theory.
Given what we learned,
we do not really expect the terms $G_4$ and $A_4$ to be coupled by
another equation. That was suggested in \cite{Sag} for the particular
example considered there. However, it is not unusual to impose further
restrictions on fields when considering a particular compactification.
In \cite{KX}, we suggested that in fact the potentials $A_3$ and $A_4$
should be fundamentally independent, and only tied by the assumption
that they correspond to a joint field strength on loop space:
\beg{egg}{(G_4,G_5)=G\in H^{4}(LZ^{12},\Z).}
This is consistent with the interpretation of the $4$ and $5$-dimensional
field strength being explained by a $3$-brane moving in the loop space.
Note however that $LZ^{12}$ is no longer low-dimensional (in fact it
is infinite-dimensional), so although an $E_8$-bundle has a characteristic
class in $4$-dimensional integral cohomology, the converse is not
true. Therefore, to use index, we need to assume in addition
to \rref{egg} that
\beg{egg1}{\text{$G$ lifts to an $E_8$-bundle on $LZ^{12}$.}
}
we then have an associated adjoint bundle $\mathcal{V}$ with $G$ on
$LZ^{12}$, and the geometric action term in F-theory which is the analogue
of the elliptic index then can be written in general as
\beg{egg2}{\frac{I_{\mathcal{V}}}{2}.
} 
We should note that we do not at this point know the correct analogue
of the Rarita-Schwinger term. In \cite{KX}, a natural way to deal with this
problem was simply to cut off the $p_3$-term, which was sufficient to
prove the results stated there.


\section{Signatures and supersymmetry}

\label{susy}
As originally noted by Vafa \cite{Vafa}, F-theory should be considered
in physical signatures. One then must ask
what is the low energy limit of such theory.
Supergravities, in the conventional sense, stop at dimension
$11$. Nevertheless, when one changes the discussion in
certain ways, various higher-dimensional scenarios
become possible. The purpose of this section is
to consider
a possibly interesting new case, which is the signature
$(9,3)$, when we consider a fundamental particle whose
world volume has three time dimensions.

\vspace{3mm}
Let us start, following the discussion in \cite{KSmod},
by recalling the Clifford algebras in twelve and eleven dimensions
with various signatures. A discussion on spinors in different dimensions and 
with various signatures can be found in \cite{Kug}.
In twelve dimensions, we are interested in $(s,t)$ signatures, with $t=0,1,2,3$.
One has symplectic Majorana-Weyl spinors in dimension $(12,0)$, majorana in dimension
$(11,1)$, Majorana-Weyl in dimension $(10,2)$ and symplectic Majorana in dimension
$(9,3)$. 
For the Lorentzian case, $(11,1)$, we have Majorana spinors. In this case, one can try 
to form a supermultiplet for supergravity formed out of 320 bosons and 320 fermions, but
the gravitino and the form sectors of the structure are incompatible \cite{Cartan}. One can
then ask whether one can construct supergravity theories with other signatures in 
twelve dimensions. A general discussion on this can be found in \cite{Sez}, and a proposal
in the $(10,2)$ signature can be found in \cite{Hew} \cite{nishino}. 
There is however a difficulty with supergravity in dimension $(10,2)$ that it
contains null states which violate Lorentz covariance.

\vspace{3mm}
In the context of \cite{KSmod}, both the $(10,2)$ and $(9,3)$ signatures
played roles in our conjectures. The main point was compactification 
of these signatures on
an elliptic curve of signature $(1,1)$ and $(0,2)$ respectively, which is
conjectured to give type IIA and IIB string theory. The point is that
in the $(10,2)$ case, the use of Lorentzian torus breaks modularity
(S-duality), which
is indeed broken in type IIA. Also the fact that full spacetime Lorentzian invariance
is broken in $(10,2)$ does not create a contradiction, since such invariance
is also broken by the elliptic curve fibration. In the $(9,3)$ case,
the $(0,2)$ elliptic curve preserves modularity, which is indeed preserved
in type IIB string theory. 

\vspace{3mm}
Connections with M-theory were also proposed
in \cite{KSmod}, schematically imagining M-theory as F-theory compactified
on a circle\footnote{The exact discussion is more complicated, see \cite{KSmod}.}.
However, in that sense, the $(10,2)$ signature contains $(10,1)$ SUGRA,
so $(10,2)$ F-theory indeed appears to contain $(10,1)$-M-theory, while 
the $(9,3)$ signature scenario seems
compatible with higher signature versions of M-theory that were
found in \cite{Mstar}. In particular, the $(9,2)$ theory in eleven dimensions
could be thought of as the dimensional reduction of the $(9,3)$ theory.

\vspace{3mm}
Let us examine what kind of supergravity could be the low energy
limit of our theory in signature $(9,3)$.
In first approximation, the author used a general formula, 
which says that for $(9,3)$ we can have symplectic-Majorana spinors.
It therefore seems possible to propose a particle content for a
$N=1$ $(9,3)$-dimensional supergravity. In lightcone gauge (with
$3$ time-like coordinates), the number of helicity 
states for the gravitino should be $(9-3-1)\cdot 
2^{(9-3)/2)}=40$.
On the other hand, the graviton has $\left(\begin{array}{c}7\\2
\end{array}\right)-1=20$ degrees of freedom, and the potential associated 
with the $G_4$
field strength has $\left(\begin{array}{c}6\\3
\end{array}\right)=20$ degrees of freedom, which would
seem to give the same number of bosons 
and fermions. 

\vspace{3mm}
A more precise analysis, however, reveals a
somewhat more subtle picture. First of all, we must 
discuss in more
detail the kind of dynamics we are considering. One can consider particles
moving on worldlines in $(9,3)$-space, but this is not what we 
want. If we did that, in the super-Poincare algebra, the odd
part would consist of spinors of the $(9,3)$-Clifford algebra, 
which are Majorana, so we would get $64$ supercharges. As usual,
half of these supercharges must act trivially on a Hilbert representation,
but that still leaves $32$ supercharges in a Clifford algebra, so the
shortest supermultiplet is the spinor which has dimension $2^{16}$.
The high number of states seems to indicate that this is probably
not the right theory. (Note that in signature $(10,2)$ one gets
Majorana-Weil spinors, i.e. $32$ independent supercharges, and the
shortest supermultiplet has dimension $256$, which remains workable.)

\vspace{3mm}
However, one can argue that in a truly physical theory,
the number of timelike dimension in spacetime and world volume should
coincide: in relativity, both notions of
time are manifestations of the same entity.
From that point of view, in signatures with $k>1$ timelike dimensions,
it is  natural to work out supergravities describing the dynamics
of particles with $k$ timelike worldvolume dimensions. Assuming that
the supergravity particles will be massless, the dimensions
which are timelike from worldvolume point of view will in fact be
lightlike from the spacetime point of view. 

\vspace{3mm}
In our case, we shall therefore consider
the case when ``world-lines'' are in fact also $3$-dimensional
worldvolumes. This means that every particle possesses $3$ independent relativistic
momenta. In a spacetime $V$ with metric of signature
$(9,3)$, the momentum then can be encoded by a tensor $V\otimes V\otimes V$.
At a definite momentum $p_1\otimes p_2\otimes p_3$, a position operator $q\in V$
acts by multiplication by
\beg{epos}{\langle q,p_1\rangle\langle q,p_2\rangle \langle q,p_3\rangle.
} 
A particle is considered {\em massless} when these three momenta are
three orthogonal vectors of norm $0$ (which is the maximum allowable
number of such vectors in signature $(9,3)$). Now the three momenta
are independent quantum numbers, and span a $3$-dimensional linear
subspace $V$ of the lightcone. By \rref{epos}, 
displacement along any vector orthogonal
to any vector of $V$ 
acts trivially on the particle. Considering a $3$-element
basis $v_1,v_2,v_3$ of $V$, considering each of the vectors $v_i$
successively cuts the space of supercharges which are allowed to
act non-trivially by $1/2$. In the end we end up with $64/8=8$ supercharges,
which form the total spin representation of $Spin(6)$ (there are two
complex conjugate
$4$-dimensional half-spin representations, which however do not possess
a real structure). Thus,
the shortest supermultiplet
is therefore just has $2^4$ elements.

\vspace{3mm}
Let us now look at the shortest massless supermultiplet in more detail
(see e.g. \cite{dw} for a review). We start with irreducible spin representations
of $Spin(9,3)$, which gives two complex (not real)
representations of dimension $32$ each. As mentioned,
fixing a nonzero momentum on the lightcone,
the dimension of the representation will be cut in $1/2$ three times, and
we obtain the two $4$-dimensional
complex spin representations $\mathbf{4^+}$ and $\mathbf{4^-}$
of $Spin(6)$. We need a real representation, so the shortest supermultiplet $M$
will be the canonical Clifford module of the Clifford algebra of
$$\mathbf{4^+} \oplus\mathbf{4^-}$$
which has dimension
$$dim(M)=2^4.$$
This is in fact completely analogous to the shortest supermultiplet in dimension $(7,1)$.
As a representation of $Spin(6)$, $M$ decomposes into bosons and fermions as
$$M_b=\mathbf{1}+\mathbf{6} +\mathbf{1},$$
$$M_f=\mathbf{4^+} +\mathbf{4^-}.$$
Here $\mathbf{6}$ is the vector multiplet.
Now the supermultiplet $M$ can be tensored with a representation $V$ of 
the transverse part of the symmetry group $Spin(6)$ in search for gravitational
supermultiplets. If we restrict ourselves to particle content of
spin $\leq 2$, the only possibility is 
\beg{emul1}{V=\mathbf{6}}
or its multiples. Taking \rref{emul1}, the supermultiplet
$$V\otimes M$$
of dimension $96$, which
has the following particle content: the fermions are
\beg{emul2}{(V\otimes M)_{f}=\mathbf{4^+} +\mathbf{4^-} +\mathbf{20^+} +
\mathbf{20^-}.
}
The first two summands are spinors, the second two summands are representations
of $Spin(6)$ of highest weight $(110)$ and $(011)$ (where $(010)$ represents
$\mathbf{6}$, $(100)$ and $(001)$ represent $\mathbf{4^+}$ and $\mathbf{4^-}$,
respectively). Thus, the last two summands constitute the real gravitino
representation, as predicted by the general formula.

\vspace{3mm}
The bosonic content may be more surprising: we have
\beg{emul3}{(V\otimes M)_{b}=\mathbf{6} +\mathbf{6} +\mathbf{1} +\mathbf{15} +\mathbf{20}.}
Therefore, we have a scalar (dilaton) and two vectors, the graviton 
$\mathbf{20}$ and an antisymmetric $2$-field (or, equivalently $4$-field)
$\mathbf{15}$. This predicts a potential
$A_2$ or $A_4$, or field strength $G_5$ or $G_3$ and not $G_4$,
which seems puzzling. Examining, however, the representations
of $Spin(6)$, we find that 
\beg{ella}{\bigwedge^{3}V\cong Sym^2(V)/\R,
}
i.e. that the graviton and the potential $A_3$ transform under the
same representation. We therefore conclude that this must be
the same particle, which we might call a ``{\em gravi-gluon}''.

\vspace{3mm}
We should point out that we haven't proved directly that the supermultiplet 
\rref{emul2}, \rref{emul3} is the correct supergravity multiplet in
dimension $(9,3)$, but it is the smallest possible,
and the only one which contains only particles with spin $\leq 2$.
One can ask how it is possible for this supermultiplet to 
contain the apparently larger supermultiplet for example of type IIB
supergravity. The explanation, however, is that we are comparing
the states of different objects. In classical supergravity, we have
a particle with one light-like worldline. In the present $(9,3)$-case,
our fundamental object is
a particle with a $3$-dimensional lightlike world volume. 
Thus, this particle has additional degrees of freedom which can absorb
any finite number of states. Another way to put it is that the
momentum representation of the Poincare group is $V\otimes V\otimes V$
instead of $V$ in the case we are considering.

\section{F-theory and duality}
\label{dual}

In this section, our goal is to examine in more detail the consistency of
a $(9,3)$-signature F-theory from the point of view of dualities
as considered in \cite{Mstar}. In \cite{KX}, we examined 
T-dualities which link F-theory fibered over IIA and IIB without
regard to signatures. Let us recall that story first.

\vspace{3mm}
In fact, some of this discussion is necessary to
explain questions left over in section \ref{lag} above. For example,
we saw that in the F-theory related to type IIA, the $12$-dimensional
spacetime $Z^{12}$ is a manifold with boundary, which is a spin-cobordism 
of the spacetime of M-theory. On the other hand, in the context of
IIB string theory, we expect $Z^{12}$ to be a closed $12$-manifold
which is elliptically fibered over $X^{10}$. Yet, these theories
should be in a string duality which would lift the T-duality between
type IIA and IIB. How is that possible?

\vspace{3mm}
In \cite{KX}, we proposed a solution along the following lines:
First recall the basic fact that when IIA is considered
on a space of the form $X^{9}\times S^1$, by shrinking the $S^1$ to a point,
that coordinate disappears, but
a string wrapped around the $S^1$ becomes light, which indicates the opening
of another dimension, thus giving the T-dual IIB theory on $X^9\times S^1$.
Now it is impossible to apply such T-duality naively to higher dimensional
theories, because of lack of fundamental strings. However,
the duality may be recovered by other means. For
example, M-theory has however $2$-branes, and one compactifies
M-theory on $S^1\times S^1$ and shrinks this $2$-torus to a point, the
$2$-brane wrapped on the torus becomes light and new dimension opens
up, giving $10$-dimensional IIB-theory. 

\vspace{3mm}
Note that in our settings, where we are using generalized
cohomology for flux quantization, there is
an additional subtlety namely that we must 
take into account spin structure on type IIA and IIB spacetime. 
Consider type IIA string theory
on $X^{9}\times S^{1}_{NS}$, which is M-theory on 
\beg{efms6}{X^{9}\times S^{1}_{NS}\times S^{1}_{R}.}
Now we know, however, that this is really an approximation
of F-theory on a spin-cobordism 
$Z^{12}$ between
\rref{efms6} and $0$. But in the special case \rref{efms6}, a 
particular choice of $Z^{12}$ can be proposed, namely
\beg{efms7a}{X^9\times E^{\prime}\times S^{1}_{R}
}
where $E^{\prime}$ is a spin-cobordism from $S^{1}_{NS}$ to $0$. Then
we can consider a process under which the size of $S^{1}_{NS}$ 
shrinks to $0$ in the boundary,
while preserving the bulk. This corresponds to gluing a disk to
$E^{\prime}$. Denoting the corresponding closed surface by $E$ (which can
be an arbitrary Riemann surface, in particular an elliptic
curve), we get the corresponding bulk F-theory on
\beg{efms7}{X^9\times E\times S^{1}_{R}.
}
Thus, we obtain indeed a ``T-duality'' between the $F$-theory with boundary
$M$-theory and $F$-theory fibered on IIB spacetime in this case.

\vspace{3mm}
In \cite{KX},
we also noted that there should be another ``self-T-duality''
of the fibered $F$-theory. Consider F-theory on
\beg{efms8}{X^{9}\times \cform{\prod}{i=1}{3}S^{1}_{R}
}
which is a special case of \rref{efms7}. Then 
this theory should have a $2$-brane $M_2$ and a $3$-brane $M_3$ where
the relationship \rref{egg} becomes 
$$M_3= M_2\times S^1.$$
In particular, then, $M_2$ can be wrapped on
$$\cform{\prod}{i=2}{3}S^{1}_{R}$$
and $M_3$ on
$$\cform{\prod}{i=1}{3}S^{1}_{R}.$$
If we shrink the radius of the first copy of $S^{1}_{R}$ to $0$, then,
$M_3$ will lose a dimension, but $M_2$ will expand by the new dimension,
and we see than that the system $(M_2,M_3)$ is self-dual.

\vspace{3mm}
Now let us look at this from the point of view of signatures,
as considered in Hull \cite{Mstar}. Hull constructs IIA and IIB-like
theories as well as M-theory in a variety of signatures. Although
these theories pass a number of consistency checks, proposing
those theories and then checking their consistency is
not the main point of \cite{Mstar}. Rather, the main point is
that these theories {\em must exist} if we make one simple assumption,
namely that in a physical spacetime, the time dimension can be compact
(i.e. topologically an $S^1$). This assumption seems to be widely accepted
now, in fact many arguments are only strictly correct if the entire
spacetime manifold is compact. Given this assumption, the theories
of \cite{Mstar} are simply constructed by applying T-duality in
the time-like dimensions.

\vspace{3mm}
In particular, following \cite{Mstar}, if we take a T-dual of a 
signature $(9,1)$ type IIB
theory in the timelike dimension, we obtain a theory in signature
$(9,1)$ denoted $\text{IIA}^*$. It differs from IIA in that in the
low energy action, the signs in the RR-sector are reversed. 
Accordingly, instead of branes which are world volumes of dimension
$(2k-1,1)$ in type IIA, we have branes which are world volumes
of dimension $(2k,0)$, i.e. time instantons. 

\vspace{3mm}
Hull continues to examine the theory $\text{IIA}^*$, in particular its
strong coupling limit. He concludes that although the
strong coupling limit is $11$-dimensional, because of the sign reversal,
the additional dimension is in fact time-like, i.e. of signature $(9,2)$.
He calls this theory $\text{M}^*$. 

\vspace{3mm}
Now let us look at this theory from the point of view of \cite{KX}.
In particular, our $\text{IIA}^*$-theory is on a spacetime of the
form
\beg{est1}{X^{9}\times S^{1}_{NS,t}
}
where $X^9$ is space-like, and the subscripts $NS$, $t$ stand
for `Neveu-Schwarz'
and `timelike', respectively. Therefore, $\text{M}^*$ is on
\beg{est2}{X^9\times S^{1}_{NS,t}\times S^{1}_{R,t}.
}
Now F-theory is on a spin cobordism of the manifold \rref{est2} to $0$.
This manifold is of the form
\beg{est3}{X^9\times E^{\prime}_{t}\times S^{1}_{R,t}
}
where $E$ is a $2$-dimensional timelike (signature $(0,2)$-) cobordism
of $S^{1}_{NS,t}$ with $0$. 

\vspace{3mm}
Now let us apply the technique of \cite{KX} of shrinking the boundary of 
$E^{\prime}_{t}$ to a point (while preserving the bulk). In this limit,
we obtain a theory on
\beg{est4}{X^9\times E_{t}\times S^{1}_{R,t}
}
where $E_t$ is $E^{\prime}_{t}$ with a disk attached. It is possible
to choose $E_t$ to be any Riemann surface, in particular
\beg{est5}{E_t=S^{1}_{R,t}\times S^{1}_{R,t},
}
in which case \rref{est4} becomes
\beg{est6}{X^9\times \cform{\prod}{i=1}{3}S^{1}_{R,t}.
}
The spacetime $X^{9}\times S^{1}_{R,t}$ where $S^{1}_{R,t}$ is the
first factor \rref{est6} is now T-dual to the original spacetime
\rref{est1}, and is therefore of type IIB. Therefore, we have constructed
F-theory of signature $(9,3)$ fibered over IIB of signature $(9,1)$
by a time-like elliptic curve.

\vspace{3mm}
Let us now briefly examine branes in this setting, and try to match them
to the supergravity sources \rref{emul3}. Specifically, let us notice that
type $\text{IIA}^*$ has a $(3,0)$-signature world volume. In the loop space
$LZ^{12}$,
we obtain world volumes of
signatures 
\beg{esig1}{(3,0)}
and
\beg{esig1a}{(3,1),\;(4,0)} 
in $\text{IIA}^*$ F-theory.
Similarly, the dual world volume in $\text{M}^*$-theory
has $(5,1)$ signature (because
$\text{M}^*$-theory has signature $(9,2)$), so in the loop space $LZ^{12}$, we 
get possible world volumes of signatures 
\beg{esig2}{(5,1), \;(6,1),\; (5,2).
}
After applying T-duality on timelike $S^{1}_{NS}$ to a point, 
the world volumes \rref{esig1} will produce $4$-dimensional world volumes
(picking up an additional time dimension), while the world volumes \rref{esig1a}
will produce $3$ or $5$-dimensional world volumes. The world volumes
\rref{esig2} will produce $5$, $6$, $7$ or $8$-dimensional world
volumes.
We see that these objects could match all the sources \rref{emul3},
plus two non-BPS states in dimension $7$ and $8$. 

\vspace{3mm}
Let us comment briefly why we aren't seeing the particle with $3$-dimensional
timelike worldvolume which gives the supergravity in dimension $(9,3)$ we
started out with in the first place. Note that one will generally
expect to see such fundamental particle as a low energy
approximation, but not a direct brane state: for example,
type IIA or IIB string theory SUGRA is a low energy approximation of
the corresponding string theory, yet the fundamental object
of this SUGRA is a particle (with $1$-dimensional world line), which
is an approximation but not directly a state of the theory. 
The present situation is analogous. We conjecture, on the basis of
field content comparison and possible supergravity supermultiplets,
that the low energy limit of signature $(9,3)$ F-theory is a supergravity
of a massless particle with $3$-dimensional worldvolume as described
above, but do not predict such particle to be seen directly as
a state of F-theory.

\vspace{3mm}
\section{Possible phenomenological predictions of $F$-theory
in signature $(9,3)$}
\label{phen}

\vspace{3mm}
One intriguing aspect of the scenario described above, i.e.
a $(9,3)$-dimensional supergravity based on a massless
particle with three time dimensions being the low energy limit
of signature $(9,3)$ F-theory, is that it offers
a possible new phenomenological scenario. The essential point
of this observation is the simple fact that
$(9,3)=3\cdot (3,1)$. It suggests quite a different
use of the ``excess dimensions'' of a higher-dimensional
description of the universe: we could conjecture that
each dimension of $(3,1)$-spacetime is in reality
a triplet of dimensions. This does not suggest a ``splitting'' of
dimensions of spacetime in the usual sense, where spacetime would
be a direct sum of observable spacetime and excess dimensions, but
in fact predicts that spacetime is locally a tensor product 
\beg{ent1}{\mathbb{R}^{9,3}=\mathbb{R}^{3,1}\otimes \mathbb{R}^{3}.
}
Therefore, a dimension reduction in this setting means that a dimensional
measurement in low energy physics in fact approximates a triplet of
measurements. It differs from other approaches to dimensional reductions in
the point that in the other types of compactification
(such as Kaluza-Klein), there is always a linear 
combination of dimensions
which gives $0$ when in terms of observed dimension. Those extra dimensions
then have to be explained, and the explanation always seems somewhat 
unnatural (in particular, one has to asks what physical principle
in those models breaks Lorentz invariance and ``freezes'' the
extra dimensions in place). The present model doesn't have
this problem, because it doesn't involve frozen dimensions.

\vspace{3mm}
There is another intriguing aspect of the present theory. One difficulty with
extending string theory beyond $10$ dimensions (including M theory) is the
fact that it is not clear what the fundamental object of the theory is
(since strings aren't critical in dimensions other than $10$).
In the present theory, we obtained a suggestion that 
in the low energy limit, the fundamental
object is a particle with three time-like world volume dimensions. 
As noted above,
it is very appealing to have a fundamental object whose number
of worldvolume timelike dimensions equals the number of spacetime
timelike dimensions. Arguably, this is required of a natural
physical theory, extending a basic principle of classical
relativity.

\vspace{3mm}
We need to explain why, in the infrared, 
a single measurement of dimension is a good
approximation for three measurements. This suggests
that there is indeed an approximate relation among
the three dimensions in a triplet (from an
observation point of view, this is particularly
important in the time dimensions). One mechanism
which could explain this is that the theory in fact
has a $3$-brane with one time-like dimension (corresponding
to the potential $A_4$). The presence of such stable brane
can produce the kind of confinement we need.
It might be worth noting that even here, there is a more
symmetrical way this may occur than simply splitting off
$4$ dimensions:
suppose, for example that we have
$$Z=X_4\times X_4\times X_4.
$$
Then the $3$-brane could be wrapped around the diagonal
\beg{ediag}{{\begin{array}{lll}
X_4&\subset& X_4\times X_4\times X_4,
\\
x&\mapsto& (x,x,x),
\end{array}
}}
which leaves more symmetries unbroken.

\vspace{3mm}
This new scenario is at this point only a proposal.
To verify it, one
would have to couple the theory to
matter and other phenomenological terms. 
Let us, however, in this paragraph, briefly speculate on
at least one phenomenon one may see there.
The lesson of string theory seems to be that strong coupling leads to
spacetime dimensional expansion. This was first shown by Witten 
when he discovered
that M-theory is the strong coupling limit of IIA string theory \cite{Var}. For 
gauge theories and sigma-models,
there have been suggestions of such nature (see \cite{Poly}).
Therefore, one might suspect that the $12$-dimensional expansion of $4$-space should
be observed in the strong coupling part of the standard model, which
is QCD. In other words, QCD phenomena might lead to local expansion of
dimension, or observable deviation of $A_4$ from the diagonal. Note also that
in signature $(p,q)$, the distance
behavior of interactions is a decrease with $r^{-p+q}$. For $p=9, q=3$,
this is $r^{-6}$, which seems to be enough for confinement, and 
closer to observation.

\vspace{3mm}
\section{Concluding remarks}
\label{egen}

\vspace{3mm}
The discussion of the present paper
leads to the possibility of a phenomenological scenario
which is potentially quite different from string-related
models proposed before.
This is because we are considering a different type of compactification,
or rather ``expansion of one dimension in signature $(3,1)$
into a triplet of dimensions''. This scenario is {\em only possible} in
signature $(9,3)$. To confirm the the theory we present here,
one needs more precise calculation of $(9,3)$-SUGRA dynamics, although 
we have reconstructed a substantial part of its action from
the effective low energy action of F-theory. Another
important feature is the ``realistic'' nature of
the dimensional expansion discussed here, which means
that the theory has a fundamental object which is
a particle with $3$ time-like world volume dimensions,
which is equal to the timelike dimensions in spacetime. 
Phenomenological Lagrangian
terms would have to be introduced to make more precise predictions,
which will be pursued in future work.


\end{document}